\documentclass[conference]{IEEEtran}
\IEEEoverridecommandlockouts
\usepackage{cite}
\usepackage{amsmath,amssymb,amsfonts}
\usepackage{algorithmic}
\usepackage{graphicx}
\usepackage{textcomp}
\usepackage{xcolor}
\usepackage{subfigure}
\def\BibTeX{{\rm B\kern-.05em{\sc i\kern-.025em b}\kern-.08em
    T\kern-.1667em\lower.7ex\hbox{E}\kern-.125emX}}
\begin{document}

\title{Intelligent analysis of EEG signals to assess consumer decisions: A Study on Neuromarketing}

\author{\IEEEauthorblockN{1\textsuperscript{st} Nikunj Phutela}
\IEEEauthorblockA{\textit{Department of Electronics and Communication Engineering} \\
\textit{PES University}\\
Bengaluru, India \\
nikunjphutela1999@gmail.com}
\and
\IEEEauthorblockN{2\textsuperscript{nd} Abhilash P}
\IEEEauthorblockA{\textit{Department of Electronics and Communication Engineering} \\
\textit{PES University}\\
Bengaluru, India \\
abhilash2599@gmail.com}
\and
\IEEEauthorblockN{3\textsuperscript{rd} Kaushik Sreevathsan}
\IEEEauthorblockA{\textit{Department of Electronics and Communication Engineering} \\
\textit{PES University}\\
Bengaluru, India \\
kaushik12499@gmail.com}
\and
\IEEEauthorblockN{4\textsuperscript{th} B N Krupa}
\IEEEauthorblockA{\textit{Department of Electronics and Communication Engineering} \\
\textit{PES University}\\
Bengaluru, India \\
bnkrupa@pes.edu}
}

\maketitle

\begin{abstract}
Neuromarketing is an emerging field that combines neuroscience and marketing to understand the factors that influence consumer decisions better. The study proposes a method to understand consumers' positive and negative reactions to advertisements (ads) and products by analysing electroencephalogram (EEG) signals. These signals are recorded using a low-cost single electrode headset from volunteers belonging to the ages 18-22. A detailed subject dependent (SD) and subject independent (SI) analysis was performed employing machine learning methods like Naive Bayes (NB), Support Vector Machine (SVM), k-nearest neighbour and Decision Tree and the proposed deep learning (DL) model. SVM and NB yielded an accuracy (Acc.) of 0.63 for the SD analysis. In SI analysis, SVM performed better for the advertisement, product and gender-based analysis. Furthermore, the performance of the DL model was on par with that of SVM, especially, in product and ads-based analysis.
\end{abstract}

\begin{IEEEkeywords}
BCI, EEG, Neuromarketing, Machine Learning, Deep Learning
\end{IEEEkeywords}

\section{Introduction}

Over the past few decades, Brain Computer Interface (BCI) has gained popularity in the field of research due to technological advancements in the field of human-computer interaction. The interactions have evolved from numerous lines of code to now being controlled by the end user’s thoughts. BCI helps in interpreting brain activity into digital form which can be used to control the computer. It eliminates the need of any physical contact between the end user and the computer, and it can be controlled by the brain signals. This is especially useful for patients with impaired motor functions such as paralysis [1].  Electroencephalogram (EEG) signals are used extensively in research related to brain activity. It is a non-invasive technique and is popular as it is easily available, cost-effective and has good temporal resolution. The increased understanding of brain activity has led researchers to study the cause-effect relationships that exist pertaining to an individual’s personal preferences. This application has been put to good use by the consumer commodity industry and has led to an emerging field of research called Neuromarketing. 

Neuromarketing harnesses the power of neuroscience and marketing in order to better understand consumer preferences and produce cost effective yet intricately designed advertisements (ads) to cater to their target audience thereby increasing sales. Some of the early work in Neuromarketing is highlighted in [2] and [3]. Traditional methods such as questionnaires and surveys are found to be ineffective as they are not an accurate indicator of the true feelings of a consumer since they are prone to dishonesty. This is where automatic technologies like EEG are used to better understand the inherent decision-making process.

In this study, EEG signals are recorded using the Neurosky Mindwave device which consists of a single electrode present at the frontal lobe of the brain. The subjects of varying ages and genders are asked to watch ads of certain products which vary in their design including and not restricted to background color, ratings, and animated GIFs, while the EEG is recorded simultaneously. The subjects are asked to note down their preferences of each ad in terms of ‘Like’, ‘Dislike’, ‘Buy’ and ‘Neutral’ which would serve as the ground truth for a classifier. The ‘Like’ and ‘Buy’ classes are combined into a ‘Positive Reaction’ class while the ‘Dislike’ and ‘Neutral’ classes are combined into a ‘Negative Reaction’ class, reducing it to a binary classification problem. The outcome of this study presents detailed results in terms of subject dependent and subject independent analysis in terms of a comparative study on the results obtained using various machine learning (ML) models and a custom deep learning (DL) model.

\section{Literature Survey}
The recent trends in EEG based signal preprocessing such as Independent Component Analysis (ICA), Common Average Reference (CAR), adaptive filters, different classification algorithms such as Adaptive Classifiers, Matrix and Tensor Classifiers, Transfer Learning and DL methods are reviewed in [4]. It concludes with the applications of EEG based BCI in different industries such as the medical, home automation, among others. It also includes the challenges faced such as the low strength of the BCI signals, high error rate, non-linearity among others.

The theoretical aspect of neuromarketing focusing on the decision making process of a consumer is explained and the results of a questionnaire are presented to understand the interest in the services of Neuromarketing in [5]. The paper describes the practical application of Neuromarketing and also addresses the ethical questions related to Neuromarketing.

The different classifiers to understand consumer choices are reflected in [6], [7], [8], [9], [10]. A modelling framework to understand consumer choices in terms of ‘likes’ and ‘dislikes’ by analyzing EEG signals is proposed in [6], [7]. A comparative analysis of classifiers such as Hidden Markov model (HMM), Random Forest and Support Vector Machine (SVM) was performed, where the HMM model was seen to outperform the others. The drawback of this paper is that a neutral choice has not been provided [6]. Different classifiers such as the Linear classifier, SVM, ANN, K Nearest Neighbors (KNN) Classifiers were used to perform a comparative analysis. The ANN classifier outperformed the other classifiers [7]. In [8] a comprehensive study was conducted to study the brain response to visual stimulus such as a short ad using a low-cost EEG headset. The ANN classifier and the Decision Tree (DT) classifier were compared and the ANN classifier with one hidden layer outperformed the DT algorithm. A comprehensive study was conducted to investigate customer inclinations on different categories of products, namely-Smartphones, Automobiles, Fast Food and Beverages. The conclusion obtained in this paper was that if the product is preferred, the value of Detrended Fluctuation Analysis (DFA) features of alpha waves will be high, and the DFA features of the beta waves will be low [9]. In [10] the potential of EEG spectral power for prediction of the customer’s preferences is discussed along with the interpretation of alteration of consumer preferences in shopping behavior based on the content of the ad. It has been observed that the ‘like’ preference increased the EEG power of theta band in the left frontal region while ‘dislike’ preference increased the theta band power in the right frontal region.

Various artifact removal techniques have been reviewed in [11], [12]. Different artifact removal techniques for EEG signals such as Principal Component Analysis (PCA), ICA, among others are used for artifact removal [11]. A novel method to enhance EEG signals by removing ocular artifacts using an adapted wavelet followed by adaptive filtering has outperformed the pre-existing methodology, Discrete Wavelet Transform (DWT). The proposed methodology has an average Mean Squared Error that is lesser than that of DWT, while the average Signal to Noise Ratio (SNR) value is higher [12].

In [13] a simple data augmentation method has been proposed in which Gaussian noise is added to the original EEG signal since other noises such as Poisson, Salt noise, or pepper noise, will change the features of the EEG data locally. To make sure the amplitude value of the sample is not changed, the data is generated with zero mean. Different values of standard deviation were taken. It has been seen that with data augmentation, the results improve in models such as LeNet, ResNet, etc.

In the recent years, DL methods are extensively used in BCI applications for extracting features and for solving several classification and regression problems. Certain CNN models like ResNet V2, Inception V3 and ResNet152 have been integrated to classify the epileptic states [14]. This integrated Deep Convolutional Neural Network (CNN) model outperformed all the current benchmark models. Another novel CNN based structure helped in extracting both temporal and spatial features without any preprocessing [15]. Furthermore, an Long Short Term Memory (LSTM) based approach employed to recognize different emotions from the raw EEG signals proved to be very promising [16], because of its powerful ability to learn features from the raw data directly. A DL model combining Representation Learning (RL), Temporal Convolutional Neural Network (TCNN) and Conditional Random Field (CRF) for automatic classification of sleep stages using a single channel EEG signal in [17]. A new data augmentation technique is also proposed to enhance the data. The model is evaluated using accuracy and kappa score as the metric and is seen to outperform the state of the art models. The limitation of this study is that it does not include hand crafted features that coud be extracted from EEG signals[18].

It can be seen that most papers use devices with multiple electrodes for data acquisition-which introduces both artifacts and noise, which can be eliminated by using a device with a single electrode. The usage of the dry electrode also helps increase participation from the subjects due to its user-friendly nature. The results of this study also include a separate analysis for subject dependent and subject independent, which includes product dependent and ad dependent which has not been performed in other recent works in this area. This study also includes the attention feature from the Neurosky Mindwave, which has not been utilized in other recent works. An option of neutral choice has also been provided in this study to the consumer,which enables the consumer to show that he/she is indifferent to the product shown in the advertisement, which has not been provided to the consumer in other works [6].

\section{Methodology}
The EEG signals are recorded using the Neurosky Mindwave device as subjects watch the advertisement on a laptop placed at a suitable distance. Their preferences are noted down and serve as ground truth for the classifier. The EEG signals are pre-processed and features are extracted which are then passed to ML and DL models which are described in detail in the subsequent sections, the illustration of which can be seen in Fig 1.

\graphicspath{{Images/}}
\begin{figure}[!ht]
    \centering
    \includegraphics[width=7cm]{"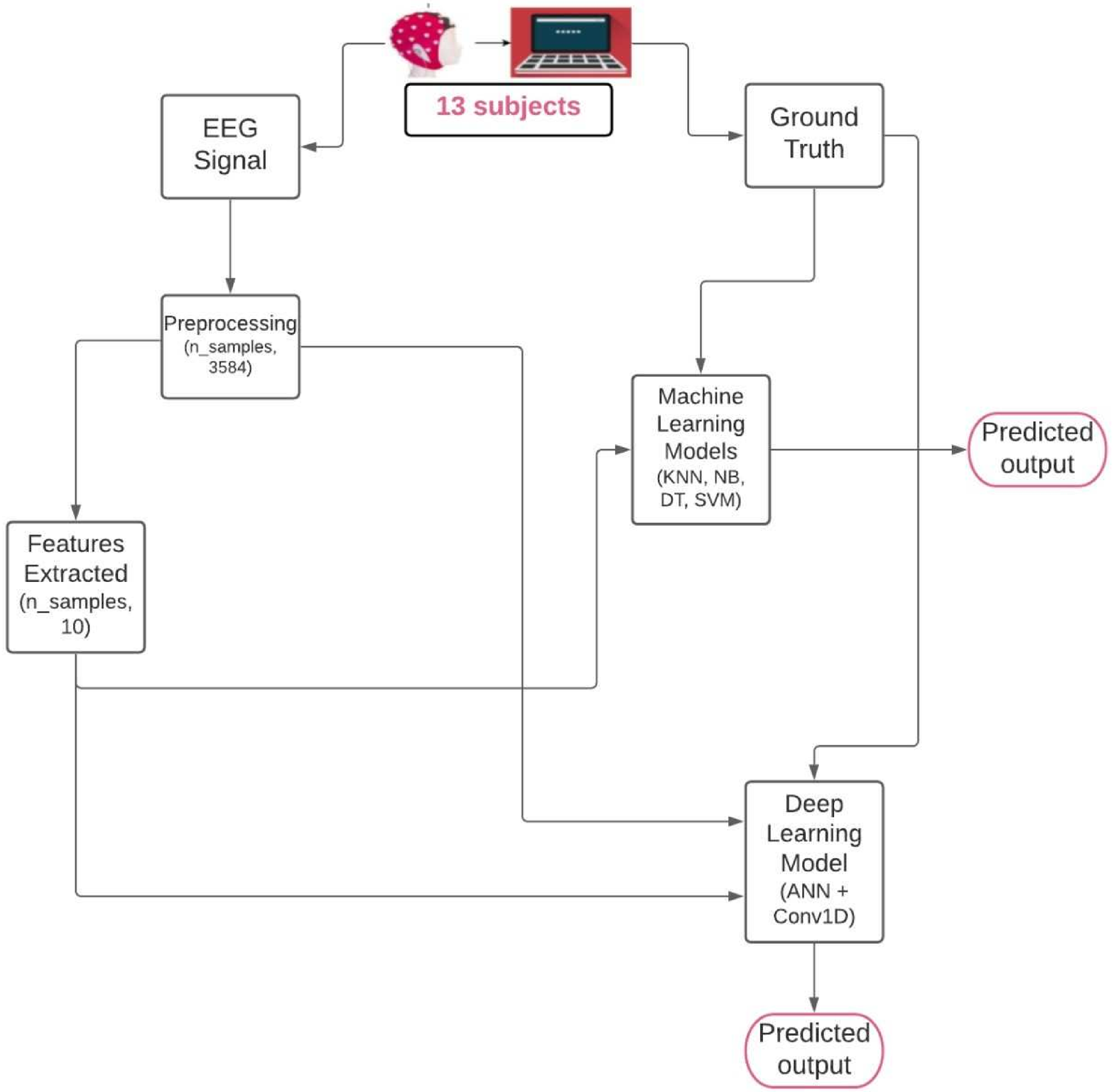"}
    \caption{Proposed methodology}
\end{figure}

\subsection{Data Acquisition}
\subsubsection{Neurosky Mindwave Electrode System} The 10-20 electrode is a standardised method for placement of EEG electrodes on the scalp, which covers all regions of the brain. It helps ensure the inter-electrode spacing is equal. As in [10], the prefrontal cortex is used for decision making while the left dorsolateral prefrontal cortex is involved in perceptual decisions and the ventromedial prefrontal cortex is used while making value-based decisions. A single electrode was used to avoid excessive noise, and to reduce the dimensionality of the data recorded. The signal can also be replicated in real time if a single electrode is used since it is a dry electrode. This has warranted the use of the Neurosky Mindwave for the data acquisition. The Neurosky Mindwave device is a headset which consists of an ear clip and a sensor arm, where the ear clip acts as the ground while the headset is being used to record raw EEG data. It consists of a reference electrode and the Fp1 electrode to record data. The Mindwave device produces the brain waves which have been sampled at 512 Hz. The output waves belong to the Alpha, Beta, Gamma, Theta and Delta spectrums. The data recorded for the purpose of analysis has been recorded with the aid of the OpenViBE software. The placement of the electrodes for the Neurosky Mindwave device is illustrated in Fig 2.

\begin{figure}[!ht]
    \centering
    \includegraphics[width=3.5cm]{"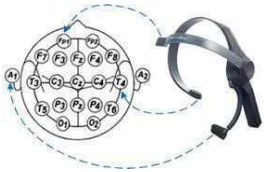"}
    \caption{Neurosky mindwave electrode system}
\end{figure}

\subsubsection{Creation of stimuli}
Five categories of products are chosen such that they don’t have a gender bias among them, namely - Sneakers, Headphones, Laptop Bags, Sunglasses and Smartphones. 
For each category a catalog of four different brands of products are selected. For example, sneakers have different brands such as Converse, North Star, Adidas and Vans. For each product, there are four different ads as mentioned in the data acquisition protocol.
\begin{itemize}
\item Ad 1 - Product with white background with no offers (as seen in Fig 3(a))
\item Ad 2 - Product with a pleasant yellow background with ratings (as seen in Fig 3(b))
\item Ad 3 - Product with a dark themed layout with ratings (as seen in Fig 3(c))
\item Ad 4 - Product with animated GIFs and offers (as seen in Fig 3(d))
\end{itemize}

\begin{figure}
    \centering
    \begin{subfigure}(a){\includegraphics[width=3.5cm]{"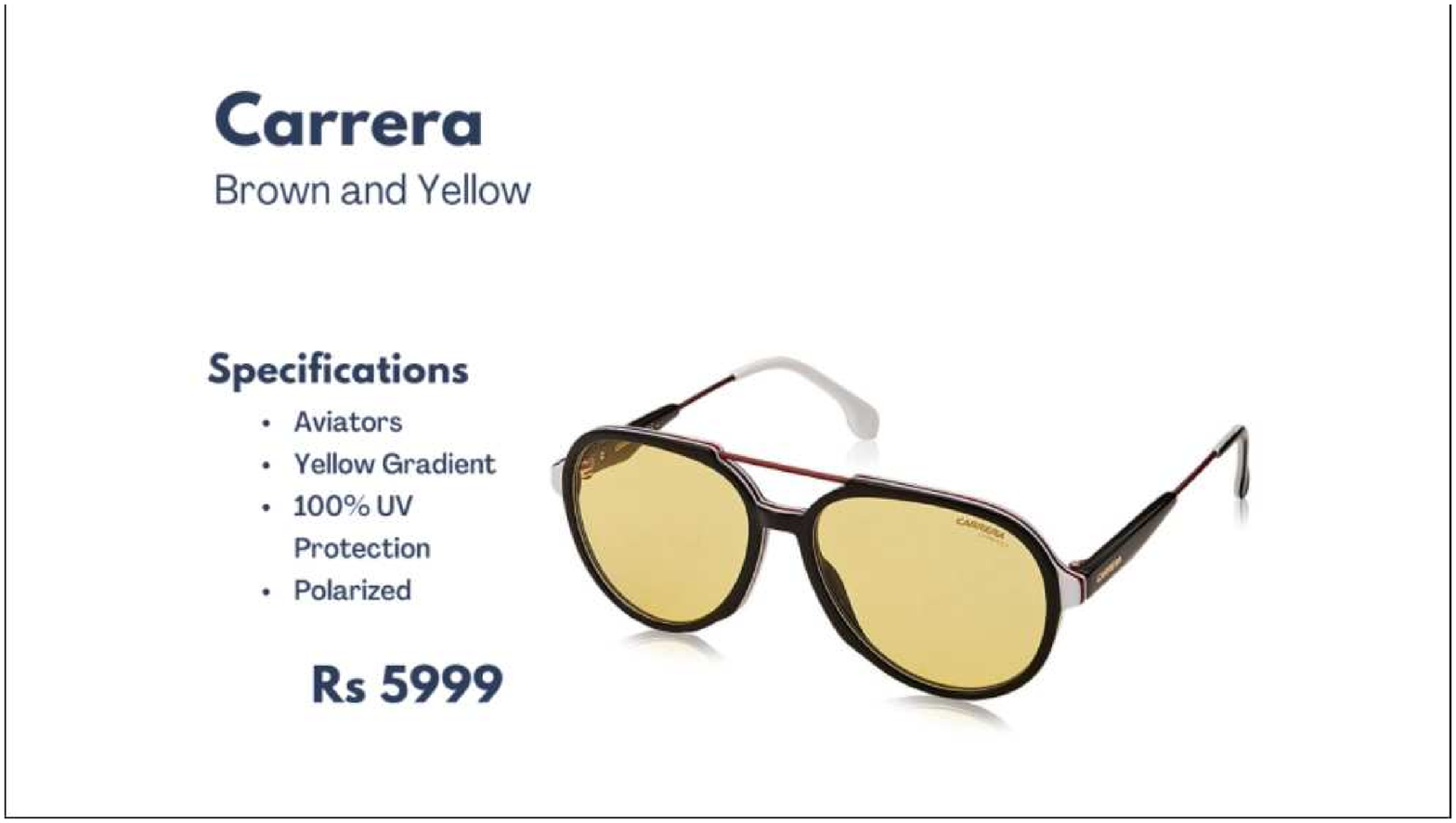"}} 
    \end{subfigure}
    \subfigure(b){\includegraphics[width=3.5cm]{"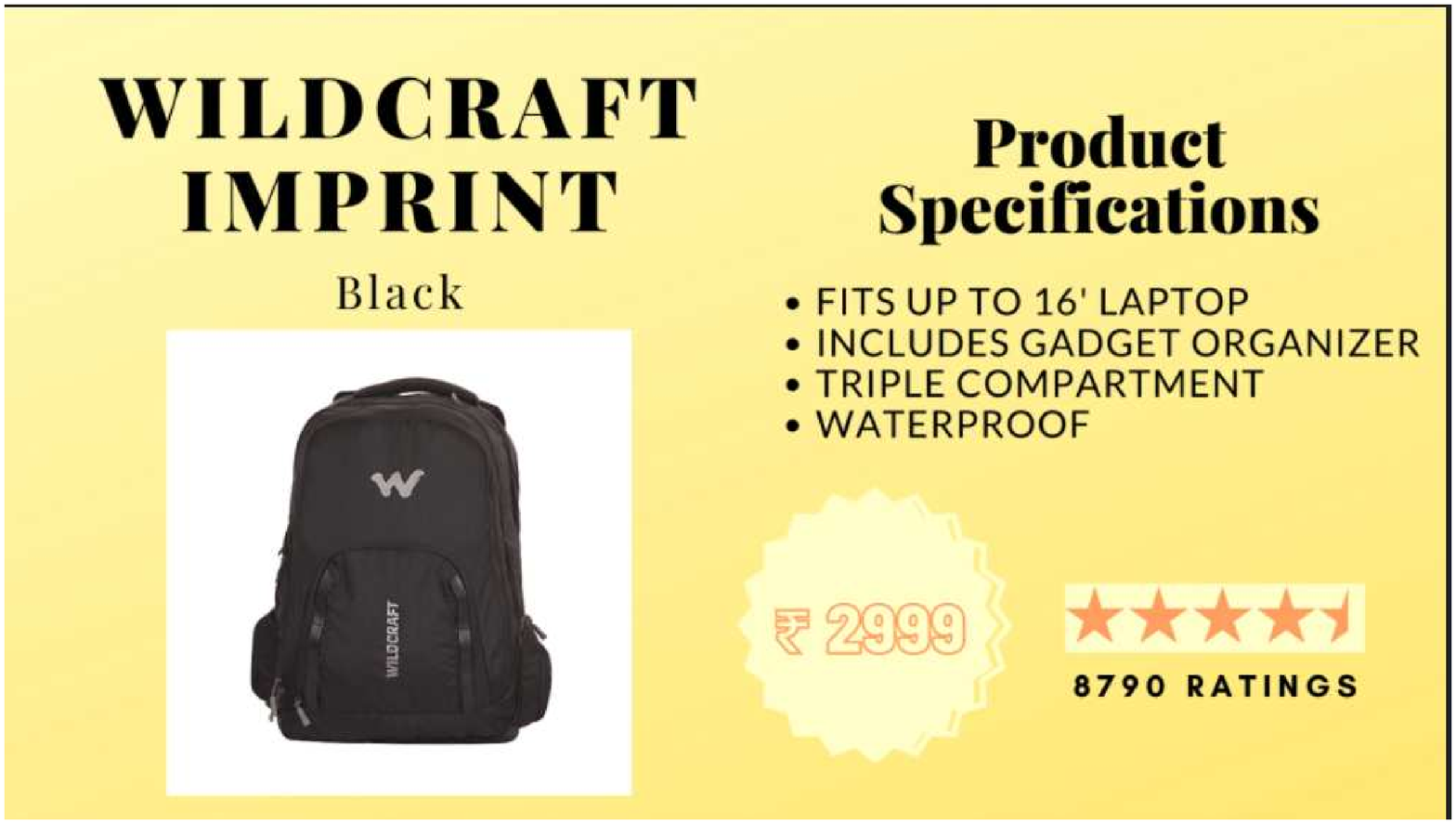"}} 
    \subfigure(c){\includegraphics[width=3.5cm]{"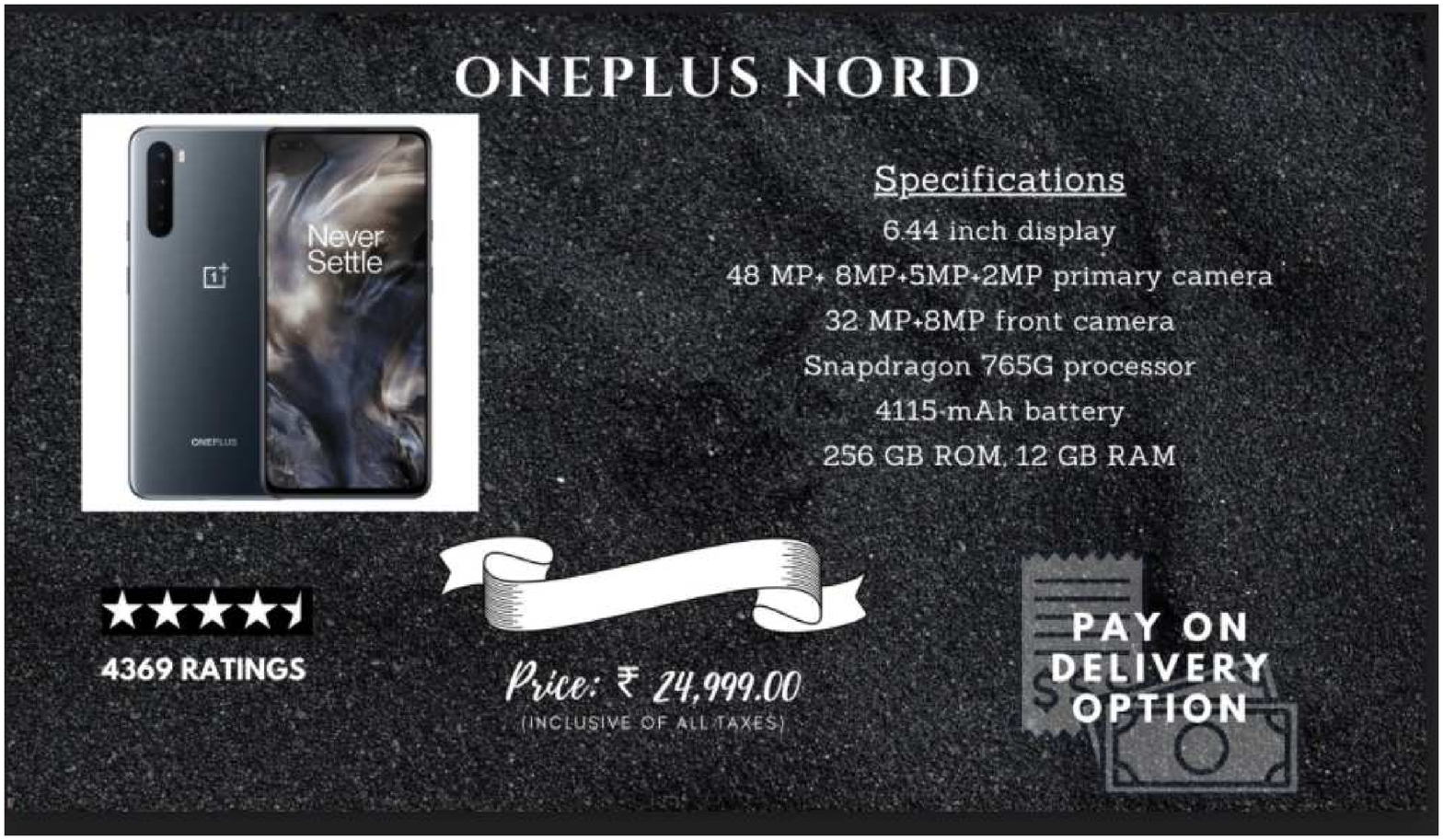"}}
    \subfigure(d){\includegraphics[width=3.5cm]{"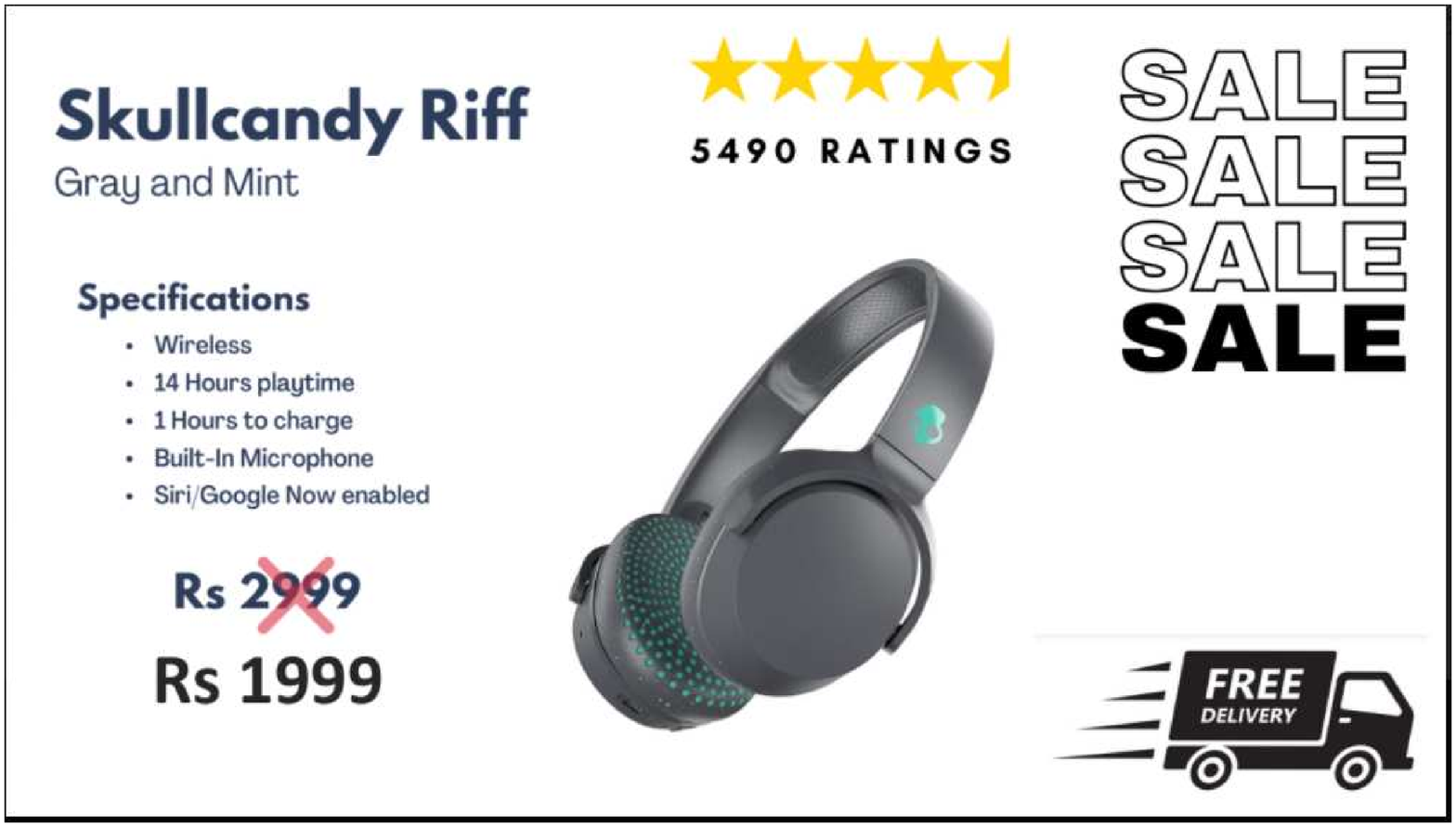"}}
    \caption{EEG signal-(a) PSD of signal before filtering, (b) PSD of signal after filtering, (c) before denoising, (d) after denoising}
    \label{fig:foobar1}
\end{figure}

As a result, there are 80 ads in total. 
(5 categories x 4 products x 4 ads)

\subsubsection{Experimental Procedure}
The study consists of 13 subjects - five males, eight females aged between 18-22.
Before undergoing the experiment, each of the subjects were asked to sign a consent form indicating their willingness to contribute to the study. Instructions were given regarding the entire process and purpose of the study. The subject was made to sit down and the Neurosky Mindwave device was placed on their head. A laptop consisting of a slideshow of the ads was placed at an appropriate distance. Subjects were told to minimize any muscular movement during the display of the ad.
Four brands of five products each were chosen and each product consisted of four different ads as already described in the previous sections constituting 80 ads in total. These 80 ads were split into eight presentations of 10 ads each. The ads were presented in a randomized order to eliminate any learning bias.
The experiment was conducted as follows:
\begin{itemize}
\item Initial 10 second resting time
\item Ad displayed for seven seconds
\item Four second buffer time between ads to note down their preference which would serve as ground truth.
\item One minute break for subjects between presentations to avoid mental fatigue.
\end{itemize}

The subjects were asked to note down their preferences as one of four categories - ‘Buy’, ‘Like’, ‘Dislike’ and ‘Neutral’ on a datasheet. Duration of the study was approximately 40 minutes per subject. The data was extracted from the Neurosky Mindwave into csv files. Each subject had 80 comma separated values (csv) files corresponding to each ad which contains 4 columns - ‘Time’, ‘Electrode’, ‘Attention’ and ‘Meditation’. 
The datasheet was collected and a separate csv file was made for each subject containing their choices in terms of ‘B’ for Buy, ‘L’ for ‘Like’, ‘D’ for ‘Dislike’ and ‘N’ for Neutral which would serve as labels for the classifier. The ‘Buy’ and ‘Like’ classes were combined into a ‘Positive Reaction’ class and the ‘Dislike’ and ‘Neutral’ classes were combined into a ‘Negative Reaction’ class, reducing it to a binary classification task. The class names indicate the response of the subject towards the ad. 

\subsubsection{Dataset Description}
The data from the Neurosky Mindwave device is sampled at 512 Hz. The study involves 13 subjects in total - five male and eight females.  In total, there are 80 ads per subject. The duration of the recording for each ad is 7 seconds and as a result 3584 samples are generated per subject per ad and the electrode values are stored in csv files along with the time stamp, attention parameter and meditation parameter extracted by the Neurosky Mindwave device. In total 1040 recordings were collected. The labels were taken from the datasheet and stored in separate csv files.

\subsection{Preprocessing}
The DC offset produces a large pulse at 0 Hz while performing the Fast Fourier Transform (FFT), which masks out the lower amplitude signals. The DC offset is eliminated by removing the mean of the signal from the original signal. An Infinite Impulse Response (IIR) Butterworth bandpass filter with lower and upper cut off frequencies at 0.5Hz and 50Hz respectively is used to remove low frequency components such as breathing and high frequency components such as noise respectively. A second order IIR notch filter at 50Hz is used to eliminate electrical line noise. The results of the Power Spectral Density (PSD) of the signal before and after filtering can be seen in Fig. 4(a) and 4(b), respectively.

An analysis of EEG signals simultaneously in the domains of time and frequency is recommended since they are highly non stationary signals. A 6-level decomposition using the ‘db7’ wavelet has been used for EEG signal denoising, as described in [17] 

\begin{figure}
    \centering
    \subfigure(a){\includegraphics[width=3.5cm]{"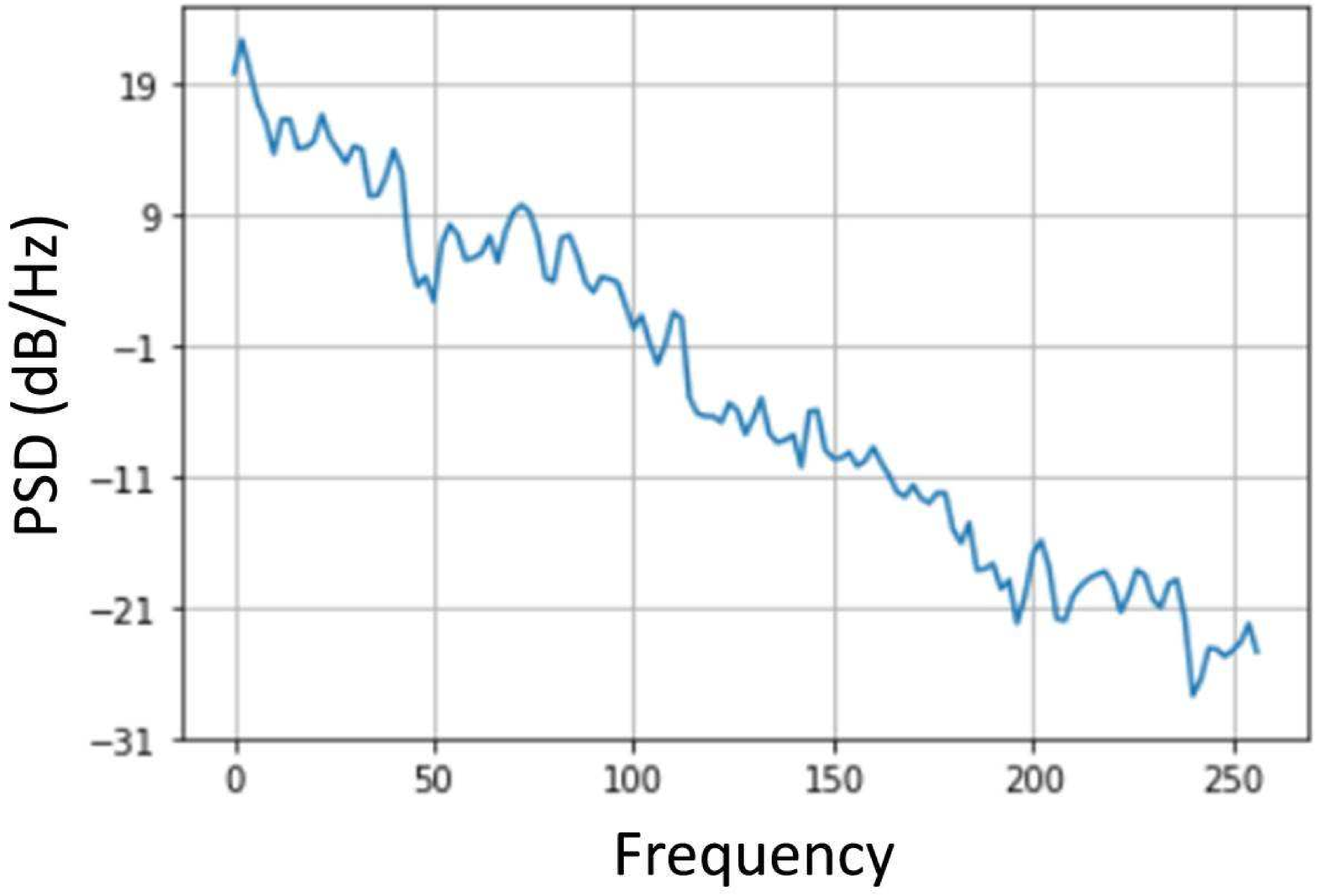"}} 
    \subfigure(b){\includegraphics[width=3.5cm]{"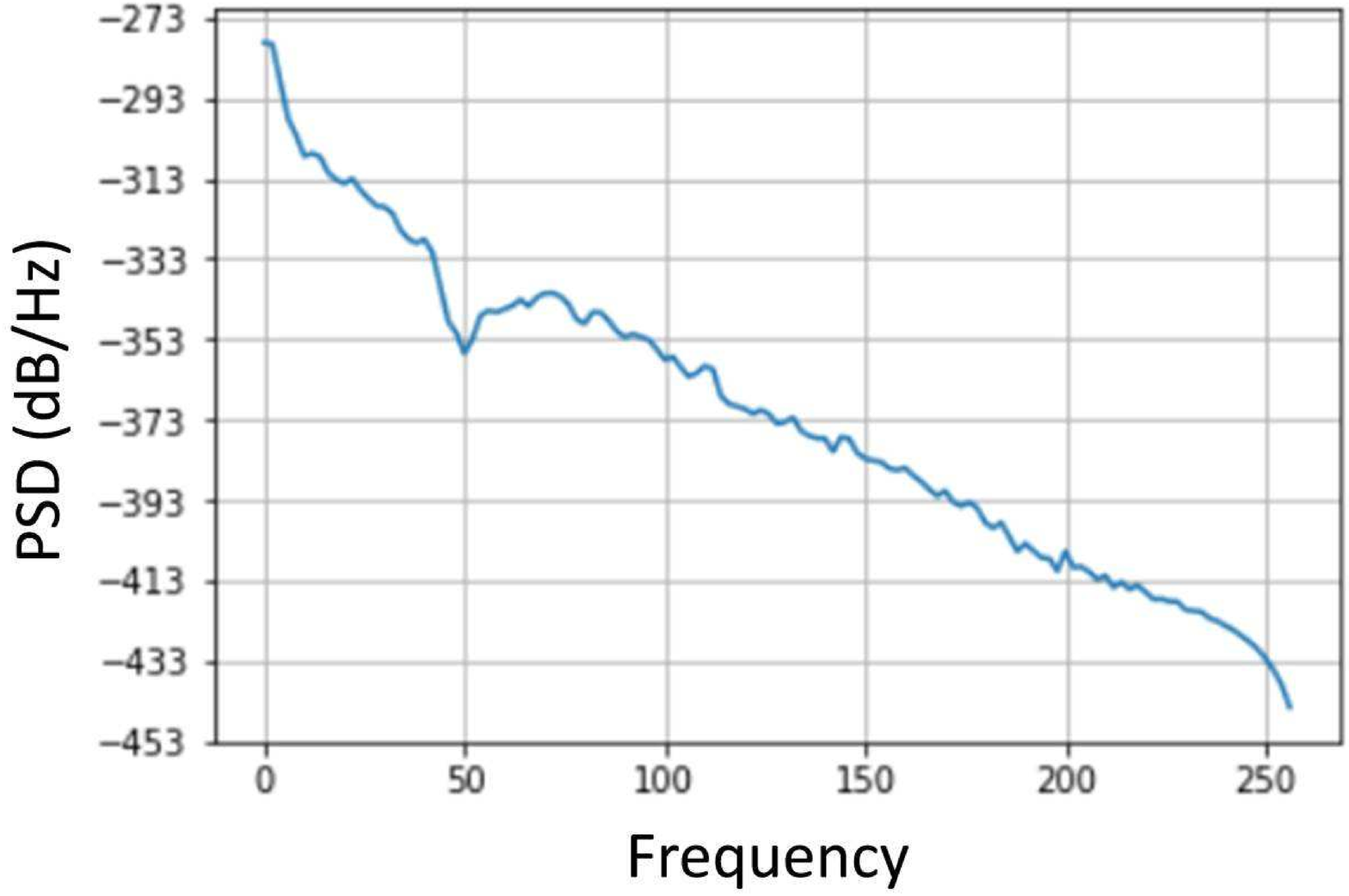"}} 
    \subfigure(c){\includegraphics[width=3.5cm]{"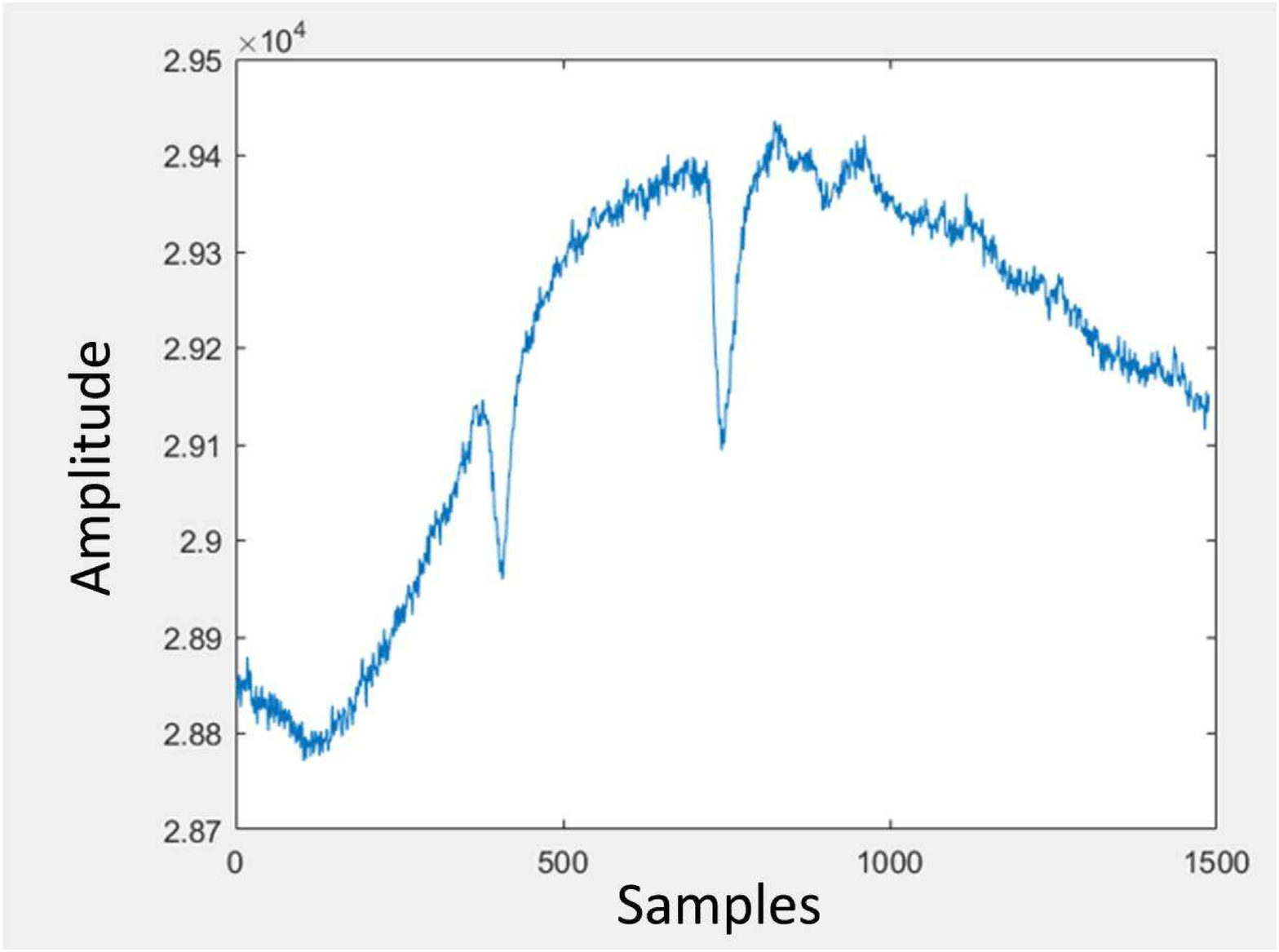"}}
    \subfigure(d){\includegraphics[width=3.5cm]{"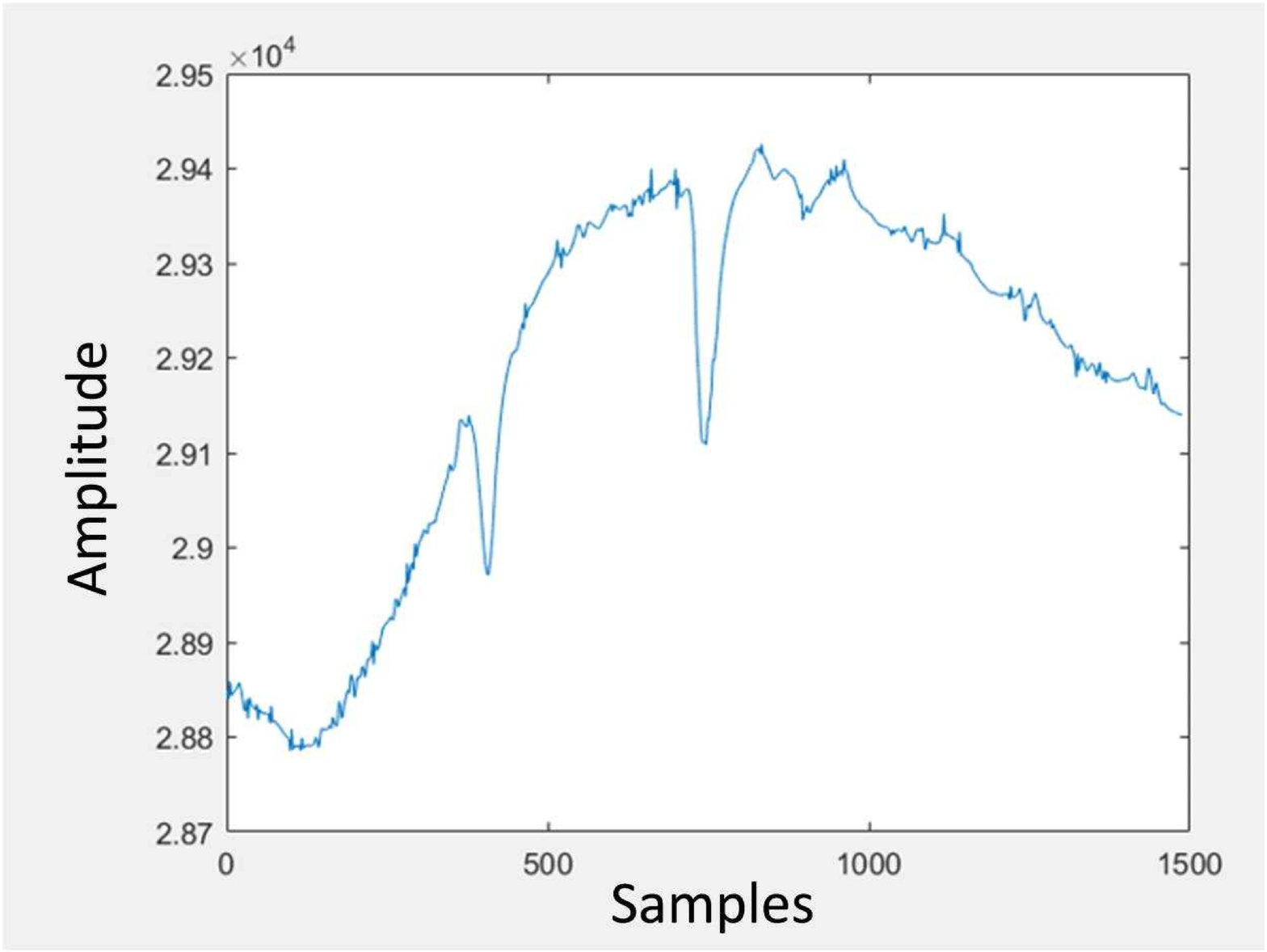"}}
    \caption{EEG signal-(a) PSD of signal before filtering, (b) PSD of signal after filtering, (c) before denoising, (d) after denoising}
    \label{fig:foobar2}
\end{figure}

Wavelet denoising uses the wavelet transform, which helps concentrate the signal and features into a few wavelet coefficients which have a large magnitude. The signal before and after denoising can be seen in Fig. 4(c) and 4(d), respectively.

\subsection{Feature Extraction}
\subsubsection{Wavelet coefficients}
The Daubechies 4 (dB4) wavelet was used to decompose the EEG signal into four levels. This decomposition provides five group wavelet coefficients D1, D2, D3, D4, and A4, where each group represents information pertaining to the five different frequency bands. The four detail coefficients - D1, D2, D3 and D4 and one approximation coefficient A4 to  represent Gamma, Beta, Alpha, Theta and Delta frequency bands respectively. The energy of each set of coefficients was calculated [6].

\subsubsection{PSD – Welch Method}
This gives an estimate of the power of the signal at various frequencies due to which a trade off occurs since noise in the power spectra reduces at the cost of reducing the frequency resolution [10].

\subsubsection{Hjorth Paramaters}
They indicate certain statistical properties used in time domain signal processing. The parameters include Mobility and Complexity, as in (1) and (2).

\begin{equation}
Mobility=\ \sqrt{\frac{variance(y\prime(t))}{variance(y(t)}}
\end{equation}

\begin{equation}
Complexity=\ \frac{Mobility(y^\prime\left(t\right))}{Mobility(y(t))}\
\end{equation}

\subsubsection{DFA}
DFA is used to determine the statistical similarity of long-memory processes like EEG signals. It can be applied to EEG signals as EEG signal statistics are non stationary [9].

\subsubsection{Attention parameter}
The attention parameter is extracted by the Neurosky Mindwave device. It indicates the magnitude of the user’s attention levels. The value lies between 0-100.  Value increases when a user focuses his attention on a particular thought while it decreases when the user is distracted. This has been extracted as  an additional feature to provide information regarding the user’s attention levels which could be influenced by the kind of ad that the user is viewing. In total 10 features are extracted per signal as shown in Table I.

\begin{table}[!ht]
\caption{Feature Vector}
\begin{center}
\begin{tabular}{|c|c|c|}
\hline
Sl No & \multicolumn{2}{c|}{Feature} \\ 
\hline
1 & {Energy of wavelet} & D1-Gamma \\ \cline{3-3} 
                   &                                    & D2-Beta    \\ \cline{3-3} 
                   &                                    & D3-Alpha   \\ \cline{3-3} 
                   &                                    & D4-Theta   \\ \cline{3-3} 
                   &                                    & A4-Delta   \\ \hline
2                  & \multicolumn{2}{c|}{Energy of power spectrum}   \\ \hline
3 & Hjorth parameter  & Mobility   \\ \cline{3-3} 
                   &                                    & Complexity \\ \hline
4                  & \multicolumn{2}{c|}{DFA}                        \\ \hline
5                  & \multicolumn{2}{c|}{Attention parameter}        \\ \hline
\end{tabular}
\label{tab21}
\end{center}
\end{table}

\subsection{Consumer Choice Classification}
This study combines ‘Like’ and ‘Buy’ classes into a single class called ‘Positive Reaction’ and ‘Neutral’ and ‘Dislike’ classes into a single class called ‘Negative Reaction’. The new class names indicate the response of the subject to the ad in terms of a positive or negative response. As a result of this, the classification task has been reduced to a binary classification problem.
The feature vector is scaled using a MinMax scaler and is then passed through the classifier. This study has incorporated ML approaches and a DL approach for the classification model.

\subsubsection{ML Models}
This study makes use of 4 ML models, namely - KNN, SVM, DT and Naïve Bayes (NB) classifier. In KNN classification the class of the sample is determined by the majority in the classes of its k-nearest neighbours. Distance is the primary metric for the computation of KNN classification.  In SVM classifiers the approach maps the training samples onto a space such that it maximises the gap between the classes and then the test samples are mapped onto the same space and the classification is obtained by determining which side of the gap they fall on. In DT, the model uses the observations made on the samples to draw conclusions on the samples target value. In the classification paradigm the leaves represent the classes and the branches stand for the conjugations of the features that lead to the classification. NB classifiers are probabilistic classifiers which are built using Bayes theorem while giving a strong prominence to the independent assumptions(naïve) between features. This model is highly scalable and also takes linear time to train when compared to the more expensive iterative approach as seen in the other classifiers. The classifiers are optimised by hyperparameter tuning.The proficiency of these models has been evidenced by [6], [7], [8], [9].

\subsubsection{DL Model}
Keras has been used to build the DL model. The features go into an Artificial Neural Network (ANN) with 3 dense layers of 30, 20 and 10 neurons while the raw signal goes into a 1D Convolutional Neural Network (CNN) with 2 convolutional layers of 128 and 32 kernels of size five followed by BatchNormalization, Dropout and MaxPooling layer. These happen parallely and then their output is concatenated and passed through a dense layers of 10 neurons and the final softmax layer for binary classification. The Conv1D is used to extract spatial information from the EEG signal and the ANN to find patterns in the features extracted from the EEG signal. The model architecture can be seen in Fig. 5.

\graphicspath{{Images/}}
\begin{figure}[htp]
    \centering
    \includegraphics[width=8cm]{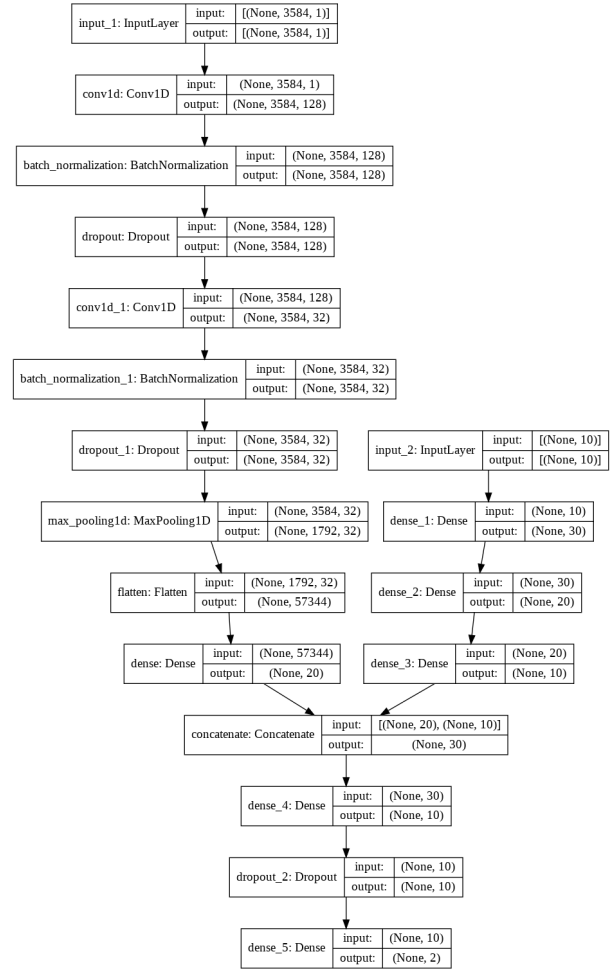}
    \caption{DL Model Architecture}
\end{figure}

\subsection{Types of analysis}
Subject Dependent analysis: The analysis here is carried out on the augmented data of the 13 subjects. \\
Subject Independent analysis: Varied types of analysis such as gender based analysis, ad based analysis and product based analysis have been performed.
\begin{itemize}
\item Gender based analysis: The analysis is carried out by splitting the recorded data according to gender. Five males and eight females are involved in the study. The data is augmented and then features are extracted for classification using the ML models.
\item Ad based analysis: The analysis is carried out by separating the recorded signals corresponding to the four different kinds of advertisements. Data is not augmented here. Each advertisement has 260 data recordings resulting in a total of 1040 recordings from which features are extracted and passed to the ML and DL models for classification based on ad type.
\item Product based analysis: The analysis is carried out by separating the recorded signals corresponding to the five different kinds of products. Data is not augmented here. Each product has 208 data recordings resulting in a total of 1040 recordings from which features are extracted and passed to the ML and DL models for classification based on product type.
\end{itemize}

\section{Results and Discussion}
The results are presented in two sections - Subject Dependent analysis and Subject Independent analysis.
\subsection{Subject Dependent Analysis}
Data augmentation has been used in order to increase the size of the dataset to help improve the accuracy (Acc.) of the models. Since EEG signals are continuous and non-stationary, they cannot be rotated or shifted as the features of the time domain will be altered. Hence Gaussian noise is added to the signal to augment the data due to its non-stationarity.
Gaussian noise with zero mean is generated to prevent the amplitude from changing and the standard deviation is varied. The Gaussian noise is then added to the signal to generate new samples. [13]
Augmented data is used in the subject dependent analysis as the number of recordings per subject is less. It has been ensured that no augmented data is present in the testing set. Each subject has 80 recordings out of which 20 random recordings are sent to the testing set and the remaining 60 are augmented six-fold and sent to the training set giving a training size of 360 samples, which are then passed to the ML and DL models. Totally 5 trials are run per model and each time a different training and testing set is created. 
The average Acc. of 10 trials is presented along with the F1 scores (F1), precision (P) and Recall (R). The results are presented in Table II. The NB and SVM classifiers perform best for subject dependent analysis with an average Acc. of 0.63 across 13 subjects. The NB model has the highest P and R of 0.54 and 0.55, respectively.

\begin{table}[!ht]
\caption{Average results of 13 subjects}
\begin{center}
\begin{tabular}{|c|c|c|c|c|c|}
\hline
Models & KNN  & SVM  & DT   & NB & DL \\ \hline
Acc. & 0.54 & 0.63 & 0.52 & 0.63 & 0.52 \\ \hline
F1 & 0.5 & 0.38 & 0.49 & 0.48 & 0.43 \\ \hline
P & 0.5 & 0.33  & 0.52  & 0.54 & 0.43 \\ \hline
R & 0.52 & 0.5 & 0.49 & 0.55 & 0.51 \\ \hline
\end{tabular}
\label{tab13}
\end{center}
\end{table}

\subsection{Subject Independent Analysis}
\subsubsection{Gender based analysis}
There are five male and eight female subjects present in the study. The average Acc. by gender across the classifiers have been presented in the Table III for males and Table IV for females. 
The NB classifier gives the highest Acc. of 0.57 across the male subjects, while both the SVM and NB produces the highest Acc. of 0.66 across the female subjects. The KNN model produces the highest P of 0.59 across the male subjects, while the KNN model produces the highest P of 0.87 across the females subjects. The NB model produces the highest R of 0.56 across the males subjects and 0.54 across the female subjects.

\begin{table}[!ht]
\caption{Average results of 5 male subjects}
\begin{center}
\begin{tabular}{|c|c|c|c|c|c|}
\hline
Models & KNN  & SVM  & DT   & NB & DL \\ \hline
Acc. & 0.52 & 0.58 & 0.52 & 0.57 & 0.54\\ \hline
F1 & 0.46 & 0.34 & 0.45 & 0.48 & 0.48 \\ \hline
P & 0.43 & 0.27 & 0.46 & 0.59 & 0.48 \\ \hline
R & 0.52 & 0.48 & 0.48 & 0.56 & 0.55 \\ \hline
\end{tabular}
\label{tab20}
\end{center}
\end{table}

\begin{table}[!ht]
\caption{Average results of 8 female subjects}
\begin{center}
\begin{tabular}{|c|c|c|c|c|c|}
\hline
Models & KNN  & SVM  & DT   & NB & DL  \\ \hline
Acc. & 0.55 & 0.66 & 0.54 & 0.66 & 0.51 \\ \hline
F1 & 0.83 & 0.69 & 0.50 & 0.48 & 0.33\\ \hline
P & 0.87 & 0.37 & 0.54  & 0.42 &  0.38\\ \hline
R & 0.53  & 0.50  & 0.5 & 0.54 & 0.49\\ \hline
\end{tabular}
\label{tab15}
\end{center}
\end{table}

\subsubsection{Ad based analysis}
Here along with the ML model, a custom DL model is used which has been described in the previous sections. There are 260 samples per ad out of which 60 go into the testing set and the remaining go into the training set. Data is not augmented here. 
The average Acc., P, R, F1  for the DL and ML models are computed across five trials.
The results are presented in terms of model Acc. and also the users reaction to each type of ad in Table V.
The SVM classifier has the highest average Acc. of 0.56 , while the DT model produces the highest P of 0.52 across the ads, while the DL model has the highest average R of 0.56 across the ads.

\begin{table}[!ht]
\caption{Average results of 4 advertisements}
\begin{center}
\begin{tabular}{|c|c|c|c|c|c|}
\hline
Models & KNN  & SVM  & DT   & NB  & DL \\ \hline
Acc. & 0.50 & 0.56 & 0.52 & 0.49 & 0.55\\ \hline
F1 & 0.51 & 0.43 & 0.52 & 0.18 & 0.5 \\ \hline
P & 0.52 & 0.44 & 0.53  & 0.50 & 0.52 \\ \hline
R & 0.50  & 0.51  & 0.53 & 0.12 & 0.56 \\ \hline
\end{tabular}
\label{tab16}
\end{center}
\end{table}

\begin{figure}[!ht]
    \centering
    \subfigure(a){\includegraphics[width=3cm]{"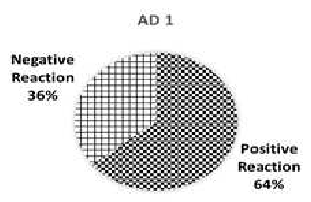"}} 
    \subfigure(b){\includegraphics[width=3cm]{"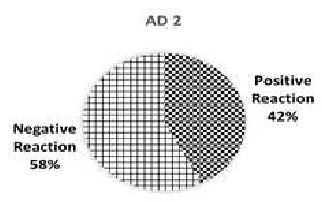"}} 
    \subfigure(c){\includegraphics[width=3cm]{"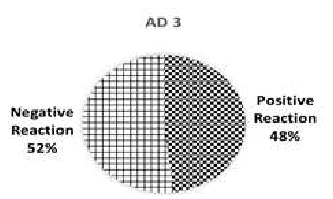"}}
    \subfigure(d){\includegraphics[width=3cm]{"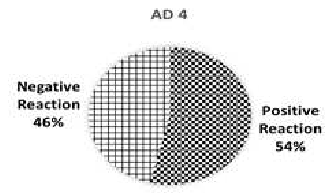"}}
    \caption{Types of Ads-(a)Ad type 1 - White background with no offers, (b) Ad type 2 - Pleasant background with ratings, (c)Ad type 3 - Dark themed with ratings, (d) Ad type 4 - Animated GIFS + offers}
    \label{fig:foobar3}
\end{figure}

It is observed from Fig. 6(a) and Fig. 6(d) that the subjects have a larger positive response towards Ad
type 1 (0.64) and Ad type 4 (0.54) while they have a larger negative response towards Ad type 2 (0.58) and Ad type 3 (0.53) as shown in Fig. 6(b) and Fig. 6(c).

\subsubsection{Product based analysis}
There are five products in the study, namely - Bags, Headphones, Phones, Sunglasses and Shoes. A total of 208 samples are present for each product where 60 go into testing data and the remaining are sent to training data. Data is not augmented here. The average Acc., F1, P, R for the DL and ML models are computed across five trials, in Table VI.
The SVM model yields the highest average Acc. of 0.58.	The NB model produces the highest P of 0.62, while the SVM and DL models produce the highest R of 0.56 across the products.

\begin{table}[!ht]
\caption{Average results of 5 products}
\begin{center}
\begin{tabular}{|c|c|c|c|c|c|}
\hline
Models & KNN  & SVM  & DT   & NB  & DL \\ \hline
Acc. & 0.50 & 0.58 & 0.51 & 0.51 & 0.56\\ \hline
F1 & 0.51 & 0.48 & 0.54 & 0.31 & 0.49 \\ \hline
P & 0.53 & 0.57  & 0.54  & 0.62 & 0.52 \\ \hline
R & 0.50  & 0.56  & 0.53 & 0.36 & 0.56 \\ \hline
\end{tabular}
\label{tab17}
\end{center}
\end{table}

The results of this paper cannot be benchmarked with similar papers in this area since there has not been a mention of subject independent and subject dependent analysis performed separately, as is done in this paper. The performance of the DL model used in this paper can be enhanced in the presence of a larger dataset. There has also been no mention of a product and ad-based analysis been performed in the recent works. But, on comparing with other papers without a mention of the different types of analysis we see that the results are similar (~60-70\%) as seen in [6], [7], [8], [9], [10].

\section{Conclusion}
In this study, EEG recordings of 13 subjects have been compiled using the Neurosky Mindwave device and both Subject Dependent and Subject Independent analysis has been performed. In the Subject Dependent analysis, the NB classifier performs best with an average Acc. of 0.61 across 13 subjects. In the Subject Independent analysis - gender based, ad based, and product-based analysis were performed and the results are analyzed. In the gender-based analysis, males have marginally better Acc. as compared to females using the SVM and KNN classifiers. In the ad-based analysis, the SVM classifier has the highest average Acc. of 0.57 across the ads, Ad type one and type four have positive reactions as compared to a negative reaction towards Ad type two and type three. In the product-based analysis, the SVM and the DL model give the highest average Acc. of 0.56. As mentioned in the results, it is difficult to compare the results of this paper with others, since no such analysis has been done in the recent works in the area.

Further additions to this project could be the inclusion of a multiclass classifier. Different algorithms to remove EEG artifacts could also be incorporated, while also including methods to assess the quality of the signals. The size of the dataset in this paper can also be enhanced, in order to explore deeper networks associated with deep learning.

\section*{Acknowledgment}

We would like to thank PES University for providing us with the resources and the assistance required during the course of this project.

\vspace{12pt}
\end{document}